%% file: B2G-17-005_temp.tex
\pdfoutput=1

\documentclass[11pt,twoside,a4paper,cmspaper,final,collab]{cms-tdr}
\def\svnVersion{468558P}\def\cmsCernNoTag{CERN-EP-2018-023}\def\cmsCernDate{\today}\def\cmsMessage{Published in the Journal of High Energy Physics as \href{http://dx.doi.org/10.1007/JHEP07(2018)075}{\doi{10.1007/JHEP07(2018)075.}}}

\begin{document}\cmsNoteHeader{B2G-17-005}

\hyphenation{had-ron-i-za-tion}
\hyphenation{cal-or-i-me-ter}
\hyphenation{de-vices}
\RCS$HeadURL: svn+ssh://svn.cern.ch/reps/tdr2/papers/B2G-17-005/trunk/B2G-17-005.tex $
\RCS$Id: B2G-17-005.tex 460680 2018-05-17 15:47:01Z plujan $

\ifthenelse{\boolean{cms@external}}{\providecommand{\cmsLeft}{left\xspace}}{\providecommand{\cmsLeft}{left\xspace}}
\ifthenelse{\boolean{cms@external}}{\providecommand{\cmsRight}{right\xspace}}{\providecommand{\cmsRight}{right\xspace}}

\newcommand{\mj}{\ensuremath{m_\text{j}}\xspace}
\newcommand{\mtvz}{\ensuremath{m_{\mathrm{T}}^{\mathrm{VZ}}}\xspace}
\newcommand{\htmiss}{\ensuremath{H_{\mathrm{T}}^{\text{miss}}}\xspace}
\newcommand{\tauTO}{\ensuremath{\tau_{21}}\xspace}
\newcommand{\ktilde}{\ensuremath{\widetilde{k}}\xspace}
\newcommand{\Mplbar}{\ensuremath{\overline{M}_{\mathrm{Pl}}}\xspace}

\cmsNoteHeader{B2G-17-005}

\title{Search for a heavy resonance decaying into a \PZ boson and a vector boson in the $\PGn\PAGn\PQq\PAQq$ final state}

\date{\today}

\abstract{
A search is presented for a heavy resonance decaying into either a pair of \PZ bosons or a \PZ boson and a {\PW} boson ($\PZ\PZ$ or \PW\PZ), with a \PZ boson decaying into a pair of neutrinos and the other boson decaying hadronically into two collimated quarks that are reconstructed as a highly energetic large-cone jet. The search is performed using the data collected with the CMS detector at the CERN LHC during 2016 in proton-proton collisions at a center-of-mass energy of 13\TeV, corresponding to a total integrated luminosity of 35.9\fbinv. No excess is observed in data with regard to background expectations. Results are interpreted in scenarios of physics beyond the standard model. Limits at 95\% confidence level on production cross sections are set at 0.9\unit{fb} (63\unit{fb}) for spin-1 \PWpr bosons, included in the heavy vector triplet model, with mass 4.0\TeV (1.0\TeV), and at 0.5\unit{fb} (40\unit{fb}) for spin-2 bulk gravitons with mass 4.0\TeV (1.0\TeV). Lower limits are set on the masses of \PWpr bosons in the context of two versions of the heavy vector triplet model of 3.1\TeV and 3.4\TeV, respectively.
}

\hypersetup{pdfauthor={CMS Collaboration},pdftitle={Search for a heavy resonance decaying into a Z boson and a vector boson in the nunubar qqbar final state},pdfsubject={CMS},pdfkeywords={CMS, physics, heavy resonances}}

\maketitle
\section{Introduction}
Many models of physics beyond the standard model (BSM) predict the existence of additional heavy resonances that may decay into a pair of vector bosons. A particular class of models addresses the divergence of quantum mechanical corrections to the Higgs boson mass, known as the hierarchy problem, by introducing extra spatial dimensions (such as warped extra dimensions models~\cite{Agashe:2007zd,Fitzpatrick:2007qr}), which predict the presence of additional massive particles.

The Randall--Sundrum model~\cite{Randall:1999ee,Randall:1999vf} introduces one warped extra dimension to solve the hierarchy problem. In the four-dimensional bulk space, two branes are hypothesized: one whose fundamental scale is the Planck scale, and one at the \TeVns scale, where the standard model (SM) particles are confined. Spin-2 gravitons, expected to have a mass at the \TeVns scale, are allowed to propagate from the Planck brane to the \TeVns brane via the warped fourth spatial dimension. In the bulk warped extra dimension model, the SM particles can also propagate through the bulk multidimensional space. In this context, spin-2 bulk gravitons can be produced at a significant rate via gluon fusion, and can decay into a pair of vector bosons~\cite{Antipin:2007pi}. Two parameters are used to describe the model: the mass of the proposed spin-2 particle and $\ktilde=k/\Mplbar$, where $k$ is the curvature parameter of the five-dimensional space-time metric, and $\Mplbar=\Mpl/\sqrt{8 \pi}$ is the reduced Planck mass.

Other theories extend the SM by adding fields to the SM Lagrangian, resulting in a larger symmetry. New vector bosons arise from the breaking of this symmetry. The heavy vector triplet (HVT) model~\cite{Pappadopulo2014} provides a framework for many BSM models, in particular those where heavy spin-1 partners of the vector bosons (\PWpr and \PZpr bosons)~\cite{Barger:1980ix,Grojean:2011vu} are expected to be weakly coupled to SM particles (referred to as the ``HVT model A'' scenario), and the composite Higgs model~\cite{Contino2011,Bellazzini:2014yua}, where exotic vector bosons are strongly coupled to ordinary particles (the ``HVT model B'' scenario). Both scenarios are described by three Lagrangian parameters: the couplings of spin-1 particles to SM fermions ($c_{\mathrm{F}}$) and to SM bosons ($c_{\mathrm{H}}$), and the strength of the interaction ($g_{\mathrm{V}}$). In the HVT model A scenario, $g_{\mathrm{V}} = 1$, $c_{\mathrm{F}} = -1.316$, and $c_{\mathrm{H}} = -0.556$; in HVT model B, $g_{\mathrm{V}} = 3$, $c_{\mathrm{F}} = 1.024$, and $c_{\mathrm{H}} = 0.976$~\cite{Pappadopulo2014}. Previous searches performed at the CERN LHC looking for evidence for these models have set limits on the production cross section of the new heavy bosons (46.1\unit{fb} at a mass of 1.4\TeV and 0.7\unit{fb} at a mass of 4.1\TeV), and mass lower limits of 3.3\TeV (3.6\TeV) for HVT model A (model B)~\cite{Sirunyan:2017acf,Sirunyan:2017wto,Aaboud:2017ahz,Khachatryan2017137,Sirunyan:2018iff}.

In this article, we present the results of a search for heavy resonances decaying into a pair of vector bosons, where one vector boson is a \PZ boson decaying into neutrinos, while the other boson V (either a {\PW} or \PZ boson) decays hadronically. The vector bosons are mostly produced in a back-to-back topology with large Lorentz boosts because of the large mass of the new particle (on the order of 1\TeV); this implies that the two quarks originating from the vector boson decay are close enough to be reconstructed within one single large-cone jet, an approach that, in this kinematic region, is more efficient than building the vector boson candidate as two distinct standard jets. Since neutrinos do not leave any visible signature in the detector, they are reconstructed as a large amount of missing transverse momentum (\ptvecmiss) recoiling against the hadronic component. The sensitivity of the search is enhanced by the relatively high branching fraction of the \PZ boson into neutrinos (20\%) and of the other vector boson into a pair of quarks ($\approx$70\%).
Jet substructure techniques~\cite{Thaler2011} are exploited to improve the discrimination between signal events and SM background processes.

The contributions of the SM backgrounds, composed mainly of \cPZ+jets and \PW+jets events, are estimated using a method that interpolates the data from control regions into the signal region with a fit constrained by the simulation.

\section{The CMS detector}
\label{CMSdetector}
The central feature of the CMS apparatus is a superconducting solenoid of 6\unit{m} internal diameter, providing a magnetic field of 3.8\unit{T}. Within the solenoid volume are a silicon pixel and strip tracker, a lead tungstate crystal electromagnetic calorimeter (ECAL), and a brass and scintillator hadron calorimeter (HCAL), each composed of a barrel and two endcap sections. Forward calorimeters extend the pseudorapidity ($\eta$) coverage provided by the barrel and endcap detectors. Muons are detected in gas-ionization chambers embedded in the steel flux-return yoke outside the solenoid.
Events of interest are selected using a two-tiered trigger system~\cite{Khachatryan:2016bia}. The first level, composed of custom hardware processors, uses information from the calorimeters and muon detectors. The second level, known as the high-level trigger (HLT), consists of a farm of processors running a version of the full event reconstruction software optimized for fast processing.

A more detailed description of the CMS detector, together with a definition of the coordinate system used and the relevant kinematic variables, can be found in Ref.~\cite{Chatrchyan:2008zzk}.

\section{Data and simulated samples}
\label{DataMC}
The analysis is performed on data collected in 2016 with the CMS detector during proton-proton collisions at the LHC at a center-of-mass energy of 13\TeV, corresponding to a total integrated luminosity of 35.9\fbinv.

Two signal models are simulated: the first considers a spin-1 HVT \PWpr boson decaying into a {\PW} and a \PZ boson for both A and B scenarios, and the second considers a spin-2 bulk graviton {\cPG} decaying into two \PZ bosons. Both processes are generated at leading order (LO) with the \MGvATNLO v2.2.2~\cite{bib:MADGRAPH} matrix element Monte Carlo (MC) generator for a range of different mass hypotheses for the resonances from 0.6 to 4.5\TeV. Signals are generated assuming the resonances have negligible width (0.1\% of their masses) compared to the experimental resolution (4--8\% depending on their masses); this assumption is the so-called ``narrow-width approximation''. The actual width of the spin-2 resonances may be larger depending on the value of the curvature parameter \ktilde in the model~\cite{Agashe:2007zd,Fitzpatrick:2007qr}, but this effect is only significant for values of \ktilde larger than 1, which are not considered in this analysis.
For the background, events with a vector boson produced with additional partons are generated at next-to-leading order (NLO) in \alpS with \MGvATNLO, using the FxFx merging scheme~\cite{bib:FXFX}.
Electroweak corrections at NLO~\cite{Kallweit:2015fta} are applied to these samples as a function of the transverse momentum \pt of the vector bosons.
Top quark-antiquark (\ttbar) and single top quark events are simulated at NLO in the five-flavor scheme with \POWHEG v2~\cite{Nason:2004rx,Frixione:2007vw,Alioli:2010xd,bib:POWHEGtt,bib:POWHEGst}.
Inclusive diboson production (\PW\PW, \PW\PZ, $\PZ\PZ$) is considered as well, and generated with \PYTHIA 8.212~\cite{bib:PYTHIA} at LO.
The hadronization and fragmentation steps of all simulated samples are handled by \PYTHIA with the CUETP8M1~\cite{bib:CUETP8M1} tune.
The NNPDF3.1~\cite{Ball:2017nwa} parton distribution functions are used in the simulations. The effect of additional proton-proton interactions within the same or nearby bunch crossings (pileup) is accounted for by adding simulated minimum bias events to the hard interaction. The frequency distributions of the pileup events are reweighted to match those observed in data.
The simulation of the CMS detector is performed with \GEANTfour~\cite{bib:GEANT4}.

\section{Event reconstruction}
\label{Obj}

The particle-flow (PF) event algorithm~\cite{Sirunyan:2017ulk} reconstructs and identifies each individual particle with an optimized combination of information from the various elements of the CMS detector. The energy of photons is obtained directly from the ECAL measurement, corrected for zero-suppression effects. The energy of electrons is determined from a combination of the electron momentum at the primary interaction vertex as determined by the tracker, the energy of the corresponding ECAL cluster, and the energy sum of all bremsstrahlung photons spatially compatible with originating from the electron track. The energy of muons is obtained from the curvature of the corresponding track.

The energy of charged hadrons is determined from a combination of their momentum measured in the tracker and the matching ECAL and HCAL energy deposits, corrected for zero-suppression effects and for the response function of the calorimeters to hadronic showers. Finally, the energy of neutral hadrons is obtained from the corresponding corrected ECAL and HCAL energies.

Jets are reconstructed from PF inputs, using \FASTJET 3.1~\cite{Cacciari:2011ma} to cluster jets with the anti-\kt algorithm~\cite{Cacciari:2008gp}, with two distance parameters: 0.4 (``AK4'' jets) and 0.8 (``AK8'' jets).
The jet momentum is determined as the four-vector sum of all particle momenta in the jet, and is found from simulation to be within 2 to 10\% of the momentum of the quark that initiated the jet, over the whole \pt spectrum and detector acceptance. The raw jet energies are further corrected to establish a relative uniform response of the calorimeter in $\eta$ and a calibrated absolute response in \pt~\cite{Chatrchyan:2011ds}. Charged particles not associated to the primary vertex are removed from the jet~\cite{CMS-PAS-JME-14-001}. An additional offset correction is applied to the jet energies to subtract the contribution from pileup~\cite{CMS-PAS-JME-14-001}. The jet energy scale (JES) is calculated using a detailed MC simulation of the detector, and further adjusted using the \pt balance in dijet, multijet, photon+jet and leptonically decaying \cPZ+jet events in data~\cite{Khachatryan:2016kdb}. A smearing procedure has been applied to jets in the simulated samples in order to account for small differences between the jet momentum resolutions observed in simulation and in data. The jet energy resolution (JER) is ${\approx}$15\% at 10\GeV, 8\% at 100\GeV, and 4\% at 1\TeV~\cite{Khachatryan:2016kdb}.

A minimum threshold on the energy recorded in the HCAL is applied to remove spurious jet-like features originating from isolated noise patterns in certain regions. Jets are required to have more than one PF constituent, and they are required to have less than 80\% of their total energy originating from neutral hadrons, less than 99\% from electrons, and more than 20\% from charged hadrons.

The jet mass reconstruction is optimized for this analysis using a combination of a jet grooming technique~\cite{Dasgupta:2013ihk,Larkoski:2014wba} and pileup mitigation~\cite{Bertolini2014}. In the jet grooming algorithm, the constituents of the AK8 jets are reclustered using the Cambridge--Aachen algorithm~\cite{Dokshitzer:1997in,Wobisch:1998wt}. The ``modified mass drop tagger'' algorithm~\cite{Dasgupta:2013ihk}, also known as the ``soft drop'' algorithm, with angular exponent $\beta = 0$, soft cutoff threshold $z_{\text{cut}} < 0.1$, and characteristic radius $R_{0} = 0.8$~\cite{Larkoski:2014wba}, is applied to remove soft, wide-angle radiation from the jet. The pileup mitigation is performed by the ``pileup per particle identification'' algorithm~\cite{Bertolini2014}, a method that assigns a weight to each charged or neutral particle, which is determined by the probability for the particle to have originated from the primary vertex of the hard interaction. Finally, the jet mass is corrected with \pt-dependent factors~\cite{Khachatryan2014} to account for the small difference observed in the reconstructed vector boson mass between data and simulated events in a \ttbar control sample, in which one {\PW} boson, originating from the top or antitop quark, decays into leptons and the other {\PW} boson decays hadronically.

The missing transverse momentum vector is defined as the negative sum of the \pt of all PF candidates in the event: $\ptvecmiss = - \Sigma_i \vec{p}_{\mathrm{T}}^{\kern1pt i}$; its magnitude is referred to as \ptmiss. This raw quantity is corrected by propagating the effect of the jet energy corrections. Uncertainties in the \ptvecmiss determination arise from mismeasurements caused by detector alignment, unclustered energy deposits, and contributions coming from pileup~\cite{CMS-PAS-JME-16-004}. Events with spurious missing momentum related to detector noise and badly reconstructed events are rejected~\cite{CMS-PAS-JME-16-004}.

\section{Event selection}
\label{Sel}

Events are required to satisfy criteria at the HLT trigger level on either \ptmiss or the missing hadronic activity, \htmiss, which is defined as the magnitude of the transverse component of the negative sum of the three-momenta of all the objects identified as jets at trigger level. To avoid inefficiencies due to the prescaling of the triggers during high-luminosity LHC operation, several triggers are used, variously requiring \htmiss or $\ptmiss > 90$, 110, 120\GeV, or $\ptmiss > 170\GeV$, in order to have at least one nonprescaled trigger at any given time.

The \ptmiss trigger efficiency has been measured with data events satisfying one or more single-muon triggers. A {\PW} leptonic decay topology is selected ($\PW \to \mu \nu$), since it ensures the presence of \ptmiss in the event, due to the neutrino. One muon identified by offline algorithms is required: this not only guarantees that the sample does not overlap with the search region of the analysis (where events with muons are rejected), but also reduces the contamination from particles or jets misidentified as leptons at the trigger level. The additional condition of having at least one AK8 jet is applied, in order to select events with a topology similar to that of the considered search. The combination of \ptmiss triggers reaches a plateau in efficiency of 96\% around $\ptmiss > 200\GeV$, which is chosen as the minimum \ptmiss threshold for the event selection. An independent efficiency measurement has been performed using a data set satisfying single-electron triggers, and the discrepancy with the result based on the muon data set is taken as a systematic uncertainty, which amounts to 1\%.

The AK8 jets are required to satisfy $\pt > 200\GeV$ and $\abs{\eta} < 2.4$. The largest-\pt AK8 jet in the event is assumed to be the hadronically decaying boson (V) candidate.

The jet mass (\mj) is used to define the search region. Since the analysis searches for a diboson resonance where one vector boson decays hadronically, the mass of the jet candidate is expected to lie within a window around the nominal masses of the {\PW} and \PZ bosons, chosen to be between 65 and 105\GeV. Two control regions are defined that are expected to be depleted in signal: the ``low sideband'', which lies in the \mj range 30--65\GeV, and the ``high sideband'', with \mj above 135\GeV. These sidebands play a crucial role in the background estimation. The region 105--135\GeV is excluded from the sideband selections in order to not overlap with other diboson searches aiming at a final state containing a hadronically decaying Higgs boson. This exclusion allows the results to be combined with those of other searches in a straightforward manner. The region under 30\GeV is discarded, since jets are not reconstructed sufficiently well in this region.

Jet substructure is exploited to further improve the ability to identify signal events.
The \tauTO $N$-subjettiness ratio~\cite{Thaler2011} distinguishes jets with two separable substructure components from jets with only one substructure component. In the former case, the \tauTO distribution is peaked towards a small fraction of unity; in the latter case, it has a broader shape, centered around larger values closer to 1. Two exclusive search categories are defined: a low-purity category ($0.35 < \tauTO < 0.75$) and a high-purity category ($\tauTO<0.35$). In principle, the high-purity category is the most sensitive to the signals explored; nevertheless, the low-purity category allows us to retain a significant part of the signal efficiency, especially for very heavy resonances (3--4\TeV). As a consequence, the signal sensitivity improves by up to 40\% when the categories are combined. Multiplicative scale factors~\cite{Khachatryan2014} are used to correct observed discrepancies between data and simulation, and are measured to be $0.99 \pm 0.11$ for events falling into the high-purity category and $1.03 \pm 0.23$ for those in the low-purity category. They have been measured with MC simulation and top quark-enriched data samples, and are applied to simulated backgrounds.

The reconstructed \ptvec of the invisibly decaying \PZ boson is set equal to \ptvecmiss. Thus, instead of the invariant mass, the resulting reconstructed V\PZ candidate mass is the transverse mass \mtvz:
\begin{equation}
\mtvz = \sqrt{\smash[b]{2 \ET^\text{j} \ptmiss \left( 1- \cos{\Delta \phi (\vec{p}^{\text{\kern1pt j}}_{\mathrm{T}},\ptvecmiss)} \right)}},
\end{equation}
where $\ET^\text{j} = E^\text{j} \sin \theta$ and $\Delta \phi(\vec{p}^{\text{\kern1pt j}}_{\mathrm{T}},\ptvecmiss)$ is the azimuthal angle between the \ptvecmiss and the leading AK8 jet transverse momentum vector.

The AK4 jets are used for background suppression; they are required to satisfy $\pt > 30\GeV$ and $\abs{\eta} < 2.4$. If the event contains an AK4 jet passing a loose b tagging criterion using the combined secondary vertex (CSVv2)~\cite{bib:btag,BTV-16-002} algorithm, and it does not overlap with the AK8 jet identified as the V candidate, the event is discarded, since this suggests that the event is more likely to have originated from a top quark decay. Scale factors are applied to correct for the different b tagging efficiency in data and simulated samples~\cite{bib:btag,BTV-16-002}.

A set of selection criteria has been applied to improve the background rejection. By requiring a minimum azimuthal angular separation of 0.5 between \ptvecmiss and the \ptvec of the AK4 jets outside the cone of the leading AK8 jet, the contribution of background events originating from soft multijet radiation is reduced from 30\% to 2\% or 3\%, depending on the purity category. The single top quark and \ttbar contributions are approximately halved by applying the loose b tag veto described above. Background contributions are further suppressed by requiring a back-to-back topology in the transverse plane between the V and \PZ candidates, specifically, $\Delta \phi > 2$.

Final states with photons, electrons, muons, and hadronically decaying tau leptons are rejected in this analysis. The identification of these objects is performed using the variables described in Ref.~\cite{Sirunyan:2017ulk}. An event is discarded if it contains at least one photon with $\pt > 15\GeV$ and $\abs{\eta} < 2.5$, at least one electron with $\pt > 10\GeV$ and $\abs{\eta} < 2.5$, at least one muon with $\pt > 10\GeV$ and $\abs{\eta} < 2.4$, or at least one hadronically decaying tau lepton with $\pt > 18\GeV$ and $\abs{\eta} < 2.4$.

\begin{figure}[!htb]
  \centering
    \includegraphics[width=.495\textwidth]{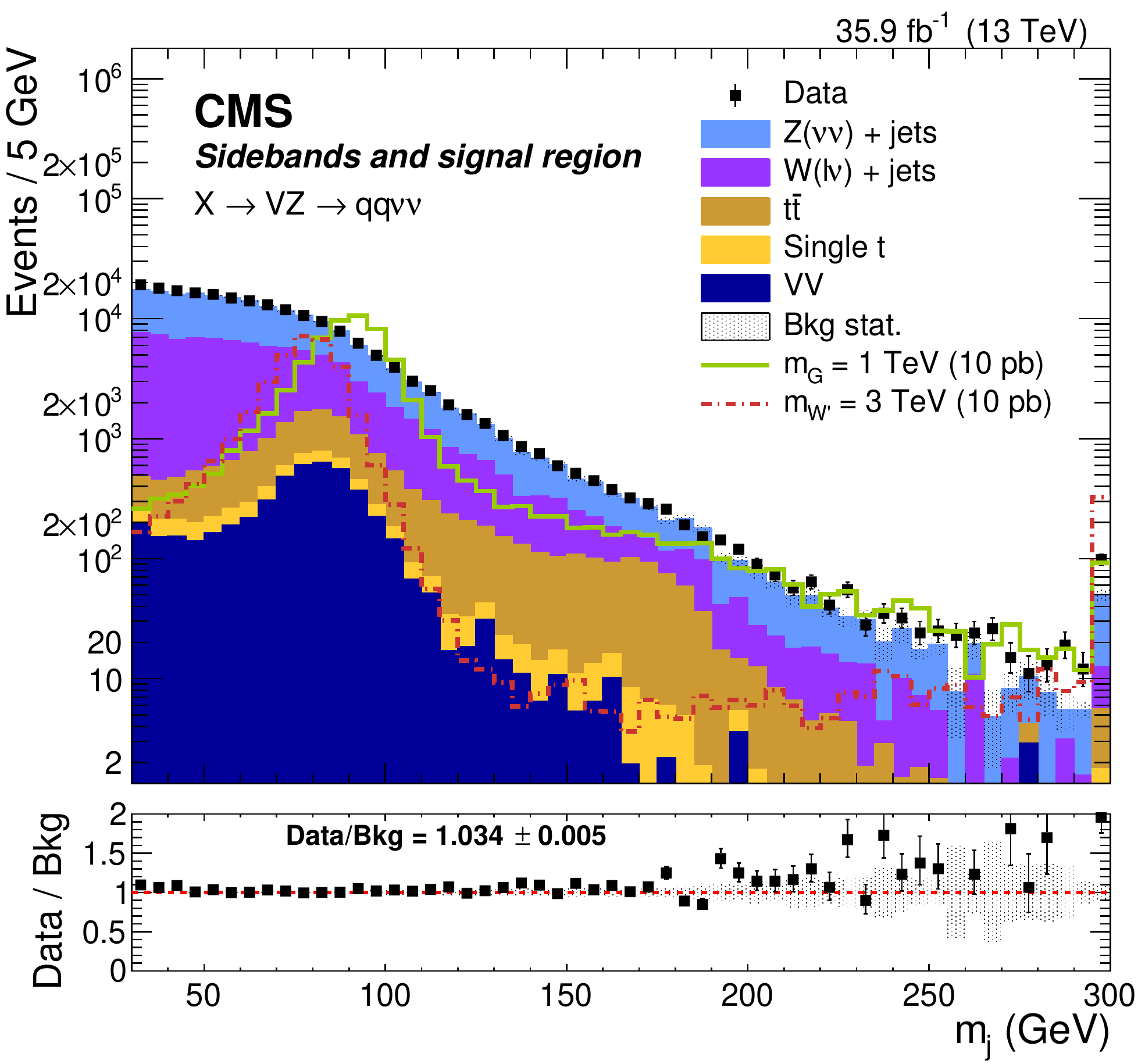}
    \includegraphics[width=.495\textwidth]{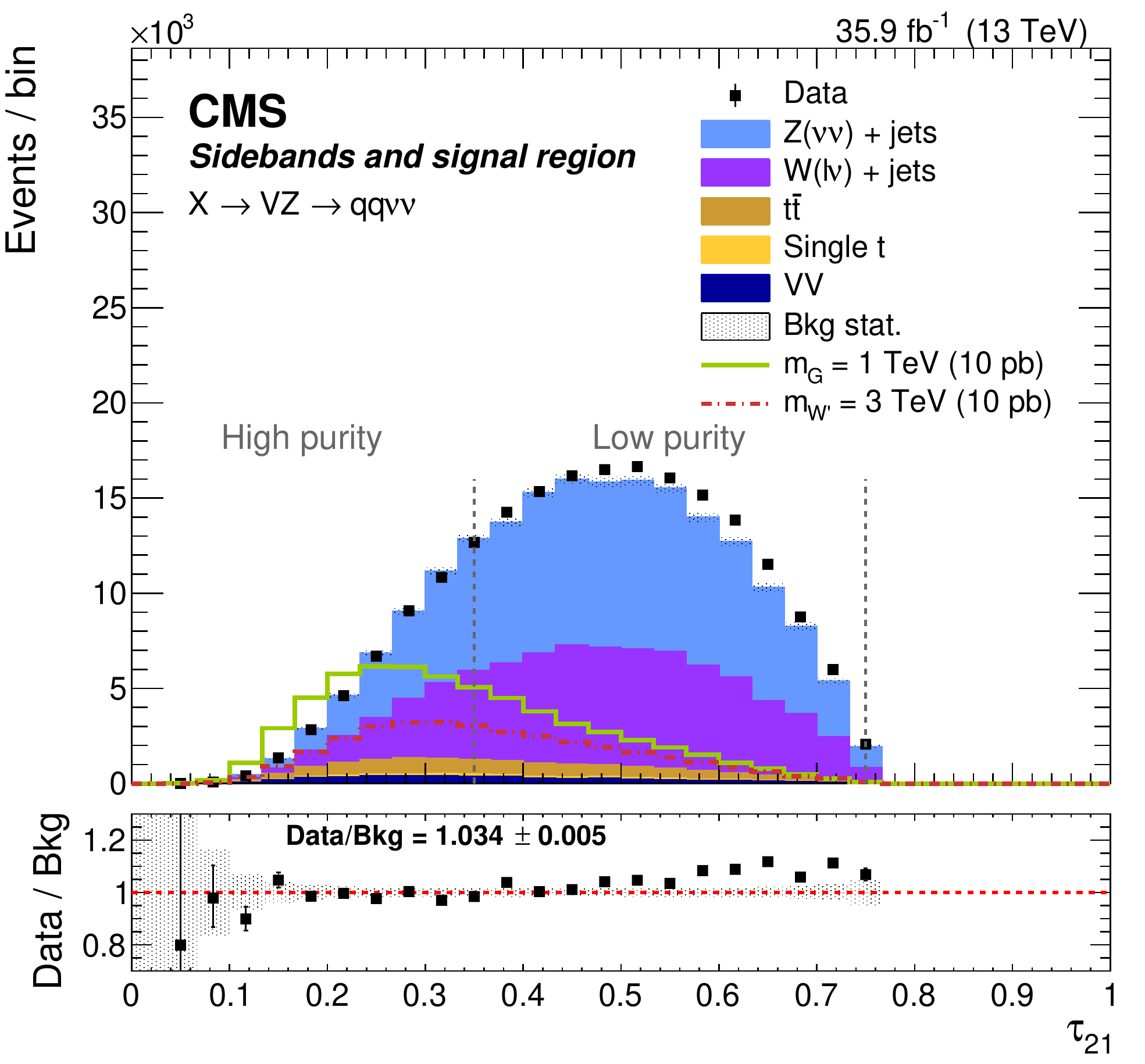}
  \caption{Comparison of data and simulated events. Left: the corrected mass of the leading AK8 jet, interpreted as the hadronically decaying vector boson. Right: the distribution of the \tauTO subjettiness of the vector boson candidate, which is used to define low- and high-purity categories.
The background processes predicted by the SM are depicted as colored filled histograms. The shaded area on top of the histograms represents the statistical uncertainty associated to MC simulations. Overflows are shown in the rightmost bin. Two possible signal hypotheses are shown: a spin-1 \PWpr boson with a mass of 3\TeV and a spin-2 bulk graviton with a mass of 1\TeV. The data points are shown by the black markers, along with their associated statistical uncertainties.
In the bottom panels, the ratio between data and MC predictions is calculated for each bin.
}
  \label{fig:XVZnn_dataMC}
\end{figure}

The main discriminating variables used to perform the background prediction, \mj and \tauTO, are compared in data and MC simulation in Fig.~\ref{fig:XVZnn_dataMC}. Two signal hypotheses, a spin-1 \PWpr boson and a spin-2 bulk graviton, are displayed as well. They are characterized by jet mass spectra peaking at the {\PW} mass and at the \PZ mass, respectively, and by a \tauTO distribution reflecting the two-prong structure of the jet produced in the vector boson hadronic decay, significantly different from the background. The discrepancy visible between the data and the background prediction is due to the imperfect modeling of the jet substructure and momentum in simulation. Agreement is achieved when a hybrid data/simulation background estimation approach, described in Section~\ref{AlphaMethod}, is applied.

\section{Background estimation}
\label{AlphaMethod}
This analysis searches for a localized excess in data in the transverse mass spectrum of the VZ system. Hence, accurate background modeling is crucial to the analysis.

The main irreducible background is from events in which a \PZ boson is produced along with additional jets (``\cPZ+jets'') and decays into neutrinos. The second dominant contribution comes from events in which a {\PW} boson is produced along with additional jets (``\PW+jets'') and decays leptonically, with the charged lepton falling outside the detector acceptance or not correctly identified. Since the production mechanisms of these two processes are the same, these two categories of events are grouped together as ``V+jets'' events. Smaller background contributions come from events in which at least one top quark (either a \ttbar pair or a single top quark, indicated as ``Top'' background) or a pair of vector bosons (\PW\PW, \PW\PZ, or $\PZ\PZ$, which we call ``VV'' background) is produced; these are referred to as ``secondary backgrounds''.

The background estimation technique~\cite{Khachatryan:2015cwa}, which is now known as the ``$\alpha$ method'', takes advantage of the data sidebands to predict the normalization and \mtvz shape of the V+jets background distributions, which are poorly populated by simulations in phase space regions with large transverse momentum. The normalization and shape of the secondary backgrounds are determined from MC simulation. This data-driven approach allows us to improve the agreement between data and predictions, especially in the higher tails of the momentum distributions.

The background prediction is performed in two steps for each of the two purity categories. First, the mass spectrum of the AK8 jet is the variable chosen to predict the background event yield in the signal region. Then, once the normalization is determined, the transverse mass distribution of the diboson candidates is used to predict the background shapes in the signal region.

To perform the normalization prediction, the \mj distribution of each background is fitted in simulated samples with an empirical probability density function (pdf), converted into an extended likelihood in order to allow the event yield to vary in the fit. The main background is modeled by using two alternative functional forms, and the difference between the two yield predictions is considered as a systematic uncertainty and propagated to the final results. The \mj spectrum of the V+jets background is smoothly falling in the low-purity category; hence, it is modeled as a power law (main function) or as a Gaussian peak added to a falling exponential (alternative function), in order to check that a different description of the slope of the spectrum near the signal region does not significantly affect the final result. In the high-purity category, the \mj spectrum has a peaking component, so it is described by a broad Gaussian peak, centered at approximately 150 GeV, added to a falling exponential (main function), or by an exponential function convolved with an error function to describe the turn-on effect at low mass (alternative function). The top quark and diboson backgrounds are modeled as Gaussian peaks, centered on the top quark and {\PW} or \PZ masses, respectively, added to a smoothly falling exponential background.

Once the extended likelihoods for the main and secondary backgrounds are added together, an extended maximum likelihood fit is performed in the data sidebands. The parameters related to the V+jets background and its normalization are allowed to vary according to data, whereas those describing the secondary backgrounds are fixed to the theoretical predictions. The expected number of background events in the signal region is then evaluated by integrating the final extended likelihood that describes the total background.

\begin{figure}[!htb]
  \centering
    \includegraphics[width=.495\textwidth]{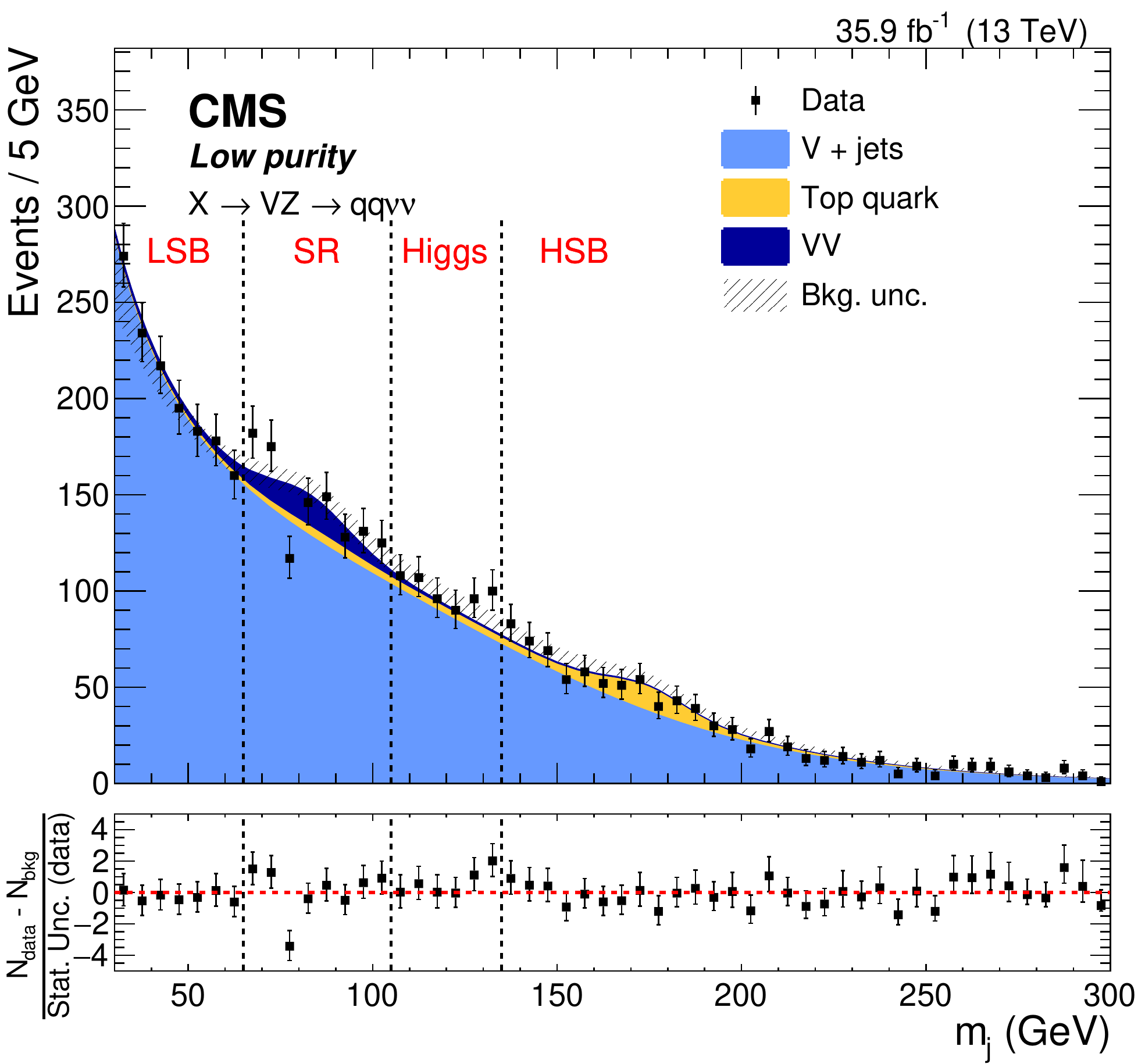}
    \includegraphics[width=.495\textwidth]{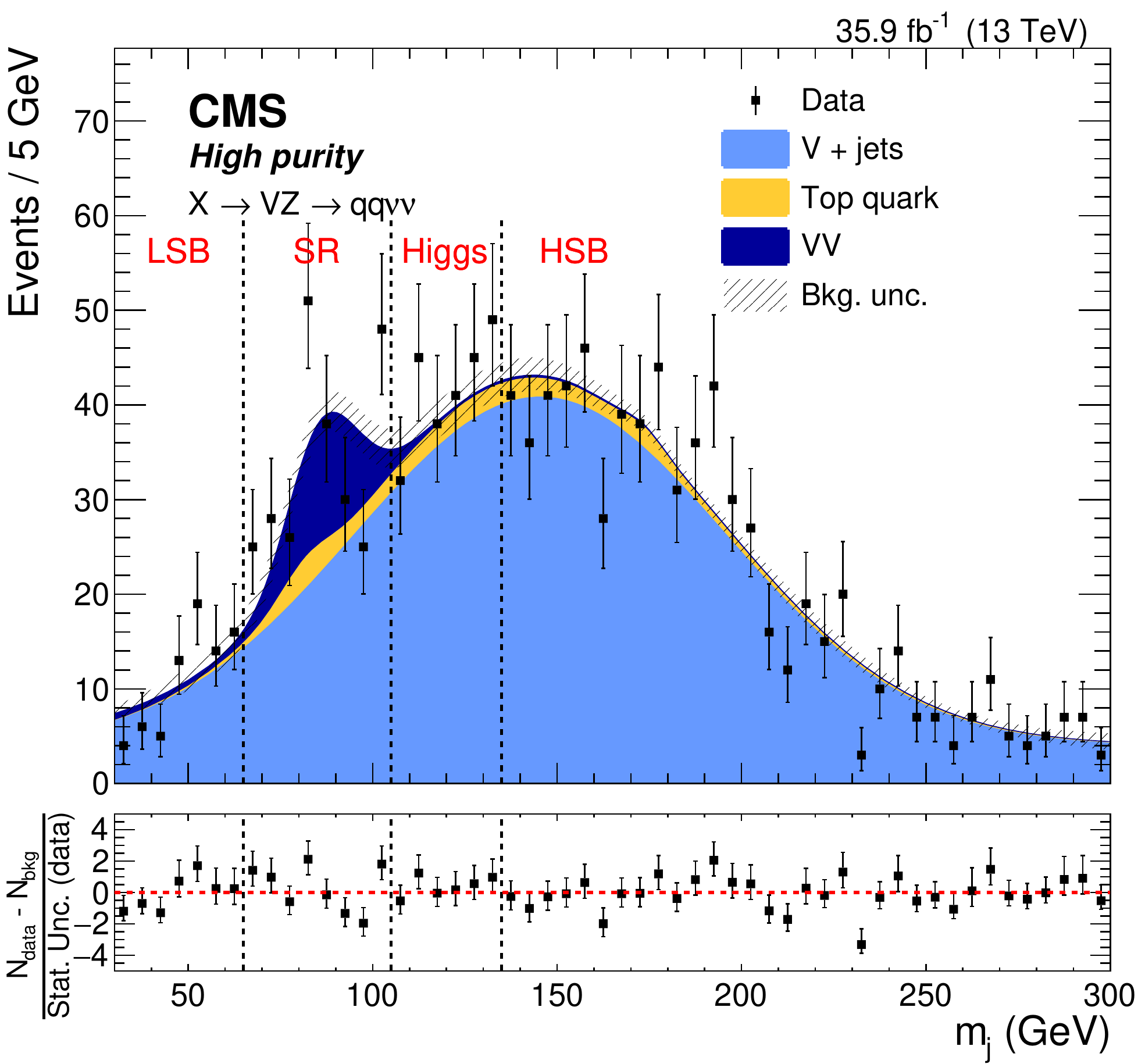}
  \caption{Background yield prediction in the signal region obtained with the $\alpha$ method, in the low-purity (\cmsLeft) and high-purity (\cmsRight) categories. Background processes predicted by the SM are depicted as colored areas bounded by smooth functions. The bottom panels show fit residuals normalized to their uncertainties.}
  \label{fig:XVZnn_JetMass}
\end{figure}

The results of the background estimation are presented in Fig.~\ref{fig:XVZnn_JetMass} as smooth functions, and are compared to data. The fit to the data is performed in the sideband regions described in Section~\ref{Sel}. Data are compared to the $\alpha$ method background predictions in the signal region (SR), while the Higgs region is excluded from the analysis. It can be seen that the data agree with the background estimates.

The final step consists in predicting the functional shape of the $m_{\mathrm{T}}$ spectrum of the total background. First, the distribution of \mtvz is described separately for each background using MC simulation, both in the signal region and sidebands. The general background shape expected for all SM processes is an exponentially falling function with two parameters, of the form $e^{-x/(a+bx)}$.

\begin{figure}[!htb]
  \centering
    \includegraphics[width=.495\textwidth]{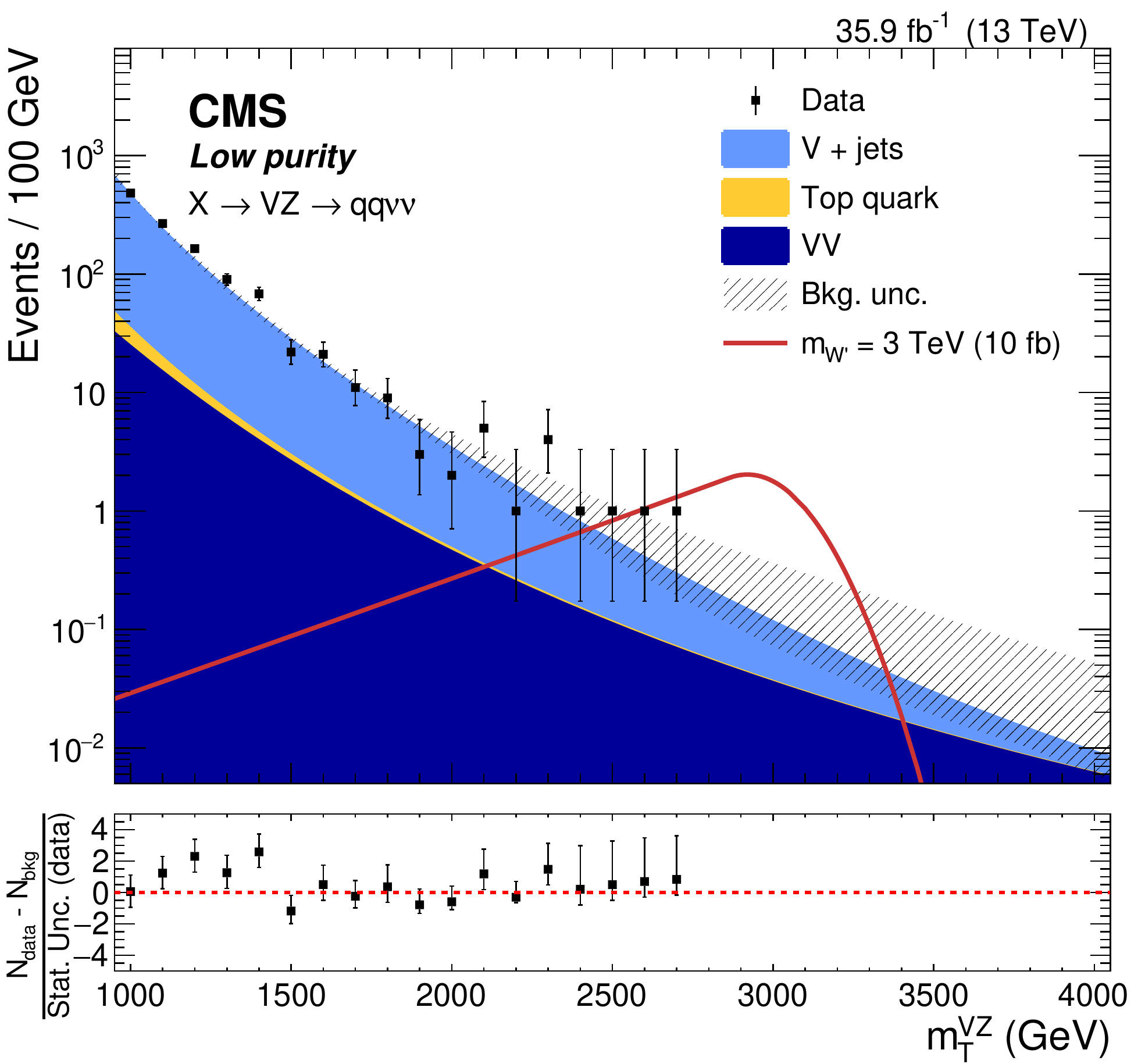}
    \includegraphics[width=.495\textwidth]{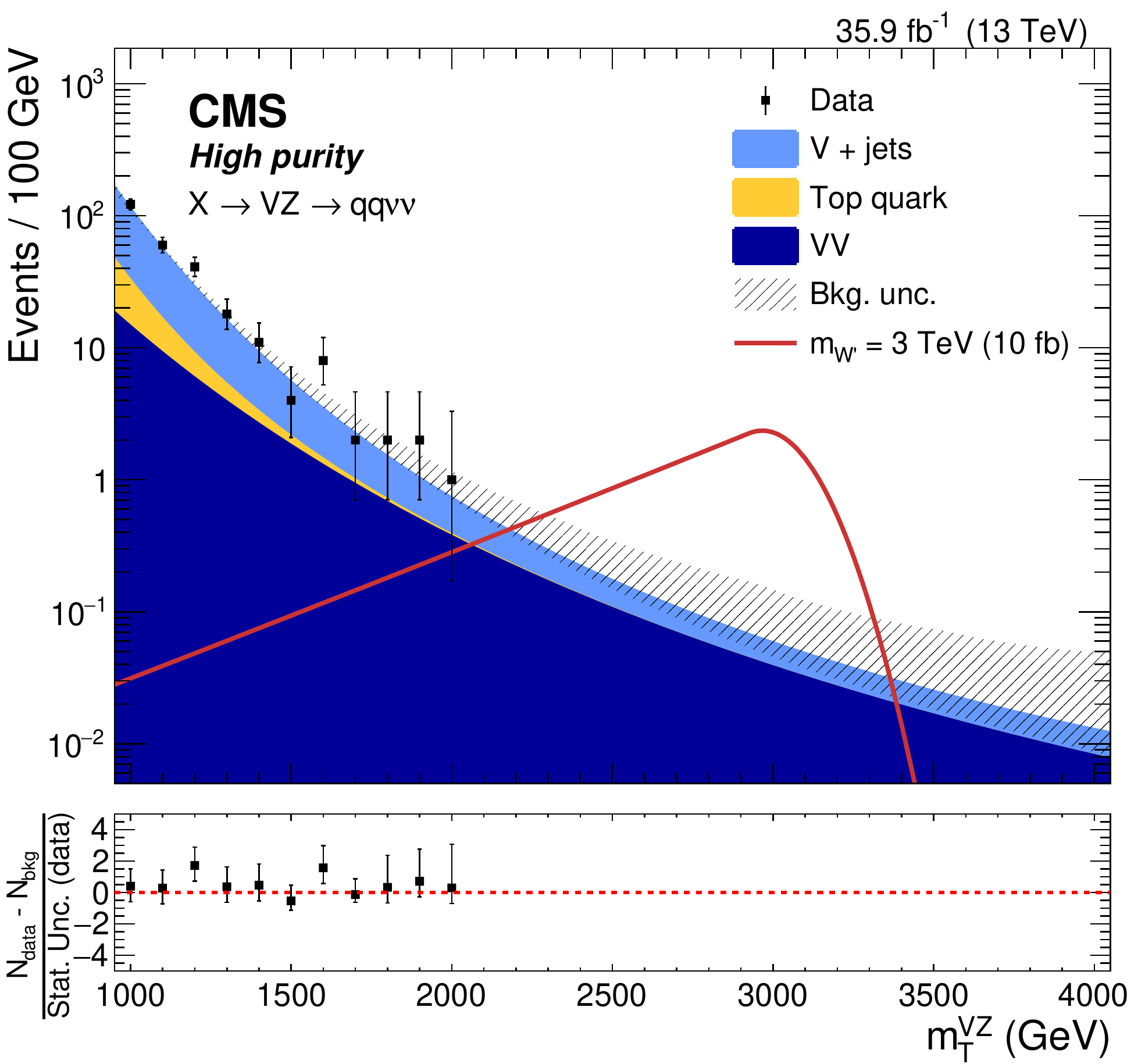}
  \caption{Expected background shapes as a function of the transverse mass of the diboson candidate obtained using the $\alpha$ method in the low-purity (\cmsLeft) and high-purity (\cmsRight) categories, represented as colored areas bounded by smooth functions.
As a reference, the expected distribution of a \PWpr with a mass of 3\TeV decaying into a {\PW} boson and a \PZ boson is displayed. Data are shown as black markers.}
  \label{fig:XVZnn_BkgSR}
\end{figure}

The $\alpha$ function is defined as the ratio between the V+jets background pdf in the signal region ($f_{\text{SR}}^{\text{V+jets}}$) and that in the sidebands ($f_{\text{SB}}^{\text{V+jets}}$), predicted from simulation:
\begin{equation}
\alpha(\mtvz) = \frac{f_{\text{SR}}^{\text{V+jets}}(\mtvz)}{f_{\text{SB}}^{\text{V+jets}}(\mtvz)}.
\label{eq:alpha_ratio}
\end{equation}
The $\alpha$ ratio can be interpreted as a transfer function from the sidebands to the signal region, accounting for the small kinematical differences in the two regions of the V+jets background. The typical correction resulting from using the $\alpha$ ratio is on the order of 1--5 per mil. A simultaneous fit to MC simulation and data sidebands is performed in order to extract the $\alpha$ function and the main background parameters respectively, while the secondary background shapes are taken from predictions from MC simulation, as described in the following equation:
\begin{equation}
f_{\text{SR}}^{\text{data}}(\mtvz) = \left[ f_{\text{SB}}^{\text{data}}(\mtvz) - f_{\text{SB}}^{\text{Top}}(\mtvz) - f_{\text{SB}}^{\text{VV}}(\mtvz) \right] \alpha(\mtvz) + f_{\text{SR}}^{\text{Top}}(\mtvz) + f_{\text{SR}}^{\text{VV}}(\mtvz).
\label{eq:shape}
\end{equation}
The background estimation obtained with the $\alpha$ method, \ie, the predicted spectrum of \mtvz in the background-only hypothesis, is compared with data in Fig.~\ref{fig:XVZnn_BkgSR}, and no significant excess is observed with regard to the SM expectations.

The robustness of the $\alpha$ method is tested by splitting the low sideband into two sub-regions, one considered as a narrower lower sideband (30--50\GeV) and the other (50--65\GeV) taken as a validation region. The predictions obtained by applying the $\alpha$ method in the narrow lower sideband and the high sideband are then compared to data distributions in the validation region, and are found to agree.

\section{Systematic uncertainties}
\label{Uncertainties}
The background normalization is predicted from a set of simultaneous fits to the simulated and data samples, so the uncertainty in the normalization is estimated by propagating all the uncertainties affecting the main and the secondary background fits. The statistical uncertainty in the fit, determined by the number of data events in the sidebands, contributes to the uncertainty in the main background event yield by 5\% and 15\% for the low- and high-purity categories, respectively. A second source of uncertainty is the absolute difference in the V+jets event yield prediction between the main function and the alternative function used to fit the \mj spectrum of the V+jets background in the simulated samples. It amounts to 5\% and 4\% for the low- and high-purity categories, respectively. The uncertainties related to the number of expected events from the secondary backgrounds amount to 68\% and 48\% for the low- and high-purity categories, respectively, for the top quark background yield, and to 11\% and 19\%, respectively, for the diboson background yield. Given that the secondary backgrounds are a small fraction of the total, the overall impact of the uncertainties in their event yields is negligible.

The uncertainties in the parameters describing the shape of the \mtvz distribution of the main background are obtained by propagating the uncertainties related to each parameter of the simultaneous fit to simulation and data sidebands. These parameters are then decorrelated by diagonalizing their covariance matrix with a linear transformation.

The normalizations of the secondary backgrounds and of the signal are affected by a 1\% uncertainty in the trigger efficiency, calculated as described in Section~\ref{Sel}.

The impact of the uncertainties in the \pt of the reconstructed bosons is evaluated by simultaneously varying their \pt within their uncertainties, since \ptvecmiss is influenced by the \pt corrections applied to all the hadronic objects present in the event. The uncertainties related to JES and JER are evaluated by varying their numerical values within their uncertainties. They have a negligible impact (less than 1\%) on both the normalization of the signal and secondary backgrounds, and on their shape; namely, on the parameters describing the exponential behavior of the spectra. The uncertainty in \ptvecmiss arising from unclustered energy deposits is also negligibly small.
Uncertainties related to the \mj corrections are considered, and they affect the signal and background yields by 1\%. Uncertainties related to the jet mass smearing affect the signal yield by 5.1\%, the top quark backgrounds by 3.1\%, and the diboson backgrounds by 2.0\%. Jet mass smearing uncertainties affect the parameters describing the top quark and diboson background shapes by 4\% and 1\%, respectively.

The uncertainty related to the \tauTO scale factors, as described in Section~\ref{Sel}, has the largest single impact on the final results. An additional source of uncertainty comes from the jet \pt dependence of the \tauTO scale factors. The \tauTO distributions are modeled at higher \pt regimes (above 200\GeV), where the event yield is very small in data, by using an alternative showering scheme (\HERWIGpp~\cite{bib:HERWIG}) and compared to \PYTHIA. The discrepancy between the predictions is parameterized as a function of the jet \pt. In this analysis, the uncertainties due to the \tauTO scale factor extrapolations at high \pt amount to 9--20\%, depending on the purity category.

The uncertainty in the b tagging efficiency affecting the veto applied to AK4 jets impacts the signal normalization by 1\%, the diboson background normalization by less than 1\%, and the top quark background normalization by 2\%.

A minor source of uncertainty comes from the uncertainty in the total inelastic proton-proton cross section at 13\TeV, which affects the pileup distribution, and thus the normalization of the simulated samples. It amounts to less than 1\% for diboson, top quark, and signal samples.

A 3\% uncertainty is assigned to the efficiency of vetoing hadronically decaying tau leptons. The uncertainty in the measurement of the integrated luminosity amounts to 2.5\%~\cite{CMS-PAS-LUM-17-001}.

The renormalization and factorization scales used in the simulation are varied by a factor of 2 and a factor of 0.5, both separately and independently. Per-event weights are extracted and propagated to the invariant mass distributions. These scale variations affect the shape of the top quark background by a total of 1\%, and its normalization by 7\% (renormalization scale) and 3\% (factorization scale); they both affect the diboson background normalization by 1\%. The uncertainty related to the choice of the parton distribution functions used in simulation is estimated by following the prescriptions in Ref.~\cite{Butterworth:2015oua}, using the NNPDF3.1~\cite{Ball:2017nwa} set. The parameters describing the parton distribution functions are varied together within their uncertainties, and the resulting variations are used as a set of per-event weights, applied to the invariant mass distributions. These uncertainties affect the normalization of the top quark and diboson backgrounds by 0.3\% each; the effect on the top quark and diboson background shapes is negligibly small. Uncertainties of 15\%~\cite{Khachatryan:2016txa,Khachatryan:2016tgp} and 10\%~\cite{Sirunyan:2016cdg,Khachatryan:2016kzg,Sirunyan:2017uhy} are assigned to the normalization of the diboson and top quark backgrounds, respectively, from the knowledge of the production cross section.

\section{Results}
\label{Results}
An unbinned profile likelihood fit is performed on the final spectra of the transverse mass of the diboson candidates. The signals are modeled with a Crystal Ball function~\cite{CBfunction}, \ie, a function with a Gaussian core and a power-law behavior in the low tail. Systematic uncertainties are treated as nuisance parameters constrained with a log-normal distribution and profiled during the minimization. The background-only hypothesis is tested in the data, where the low- and high-purity categories have been combined. The asymptotic modified frequentist approach~\cite{bib:CLS1, bib:CLS2, Cowan:2010js}, or CL$_\mathrm{s}$ criterion, is used to quote 95\% confidence level (CL) limits.

\begin{figure}[!htb]
  \centering
    \includegraphics[width=.495\textwidth]{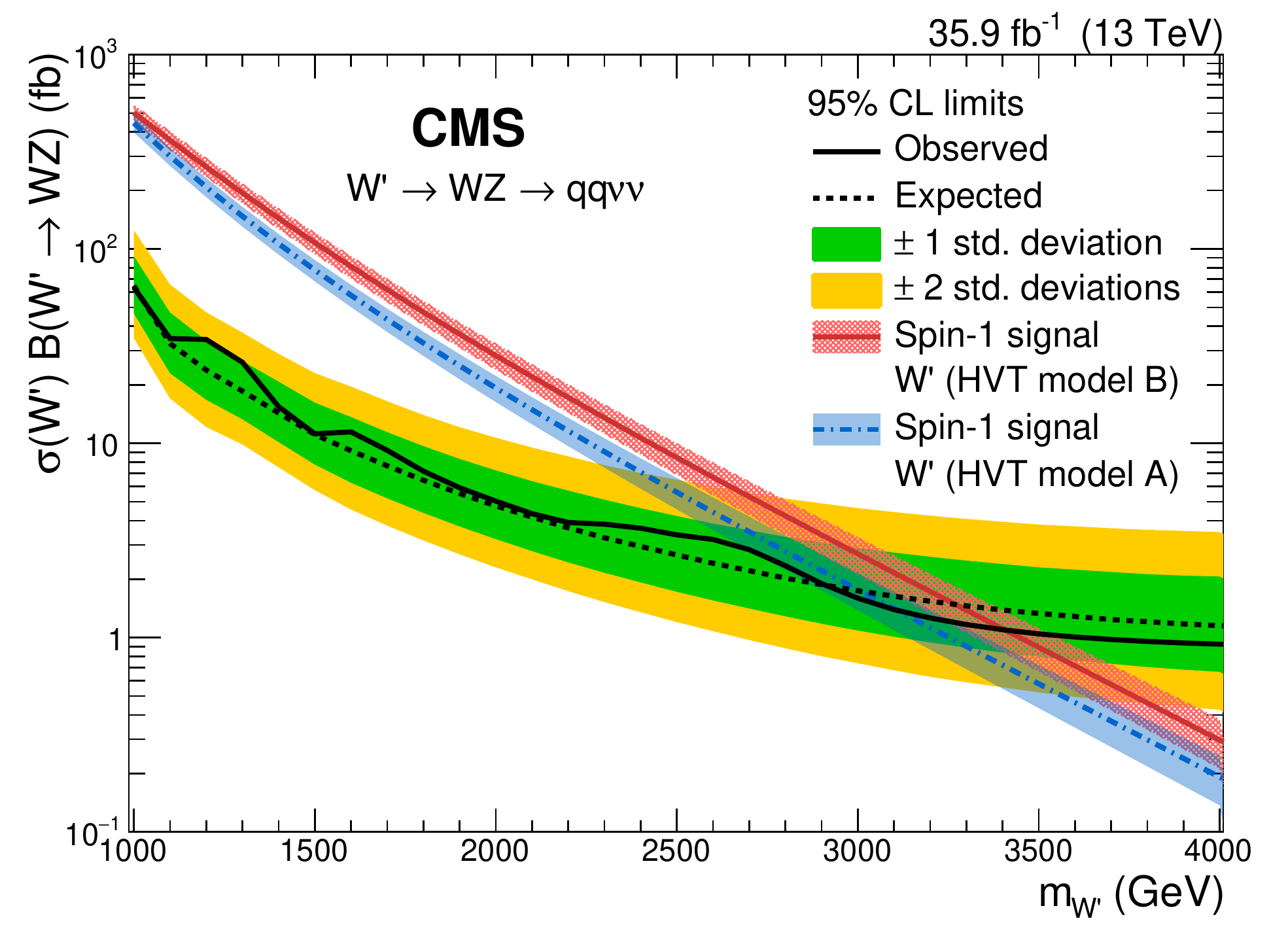}
    \includegraphics[width=.495\textwidth]{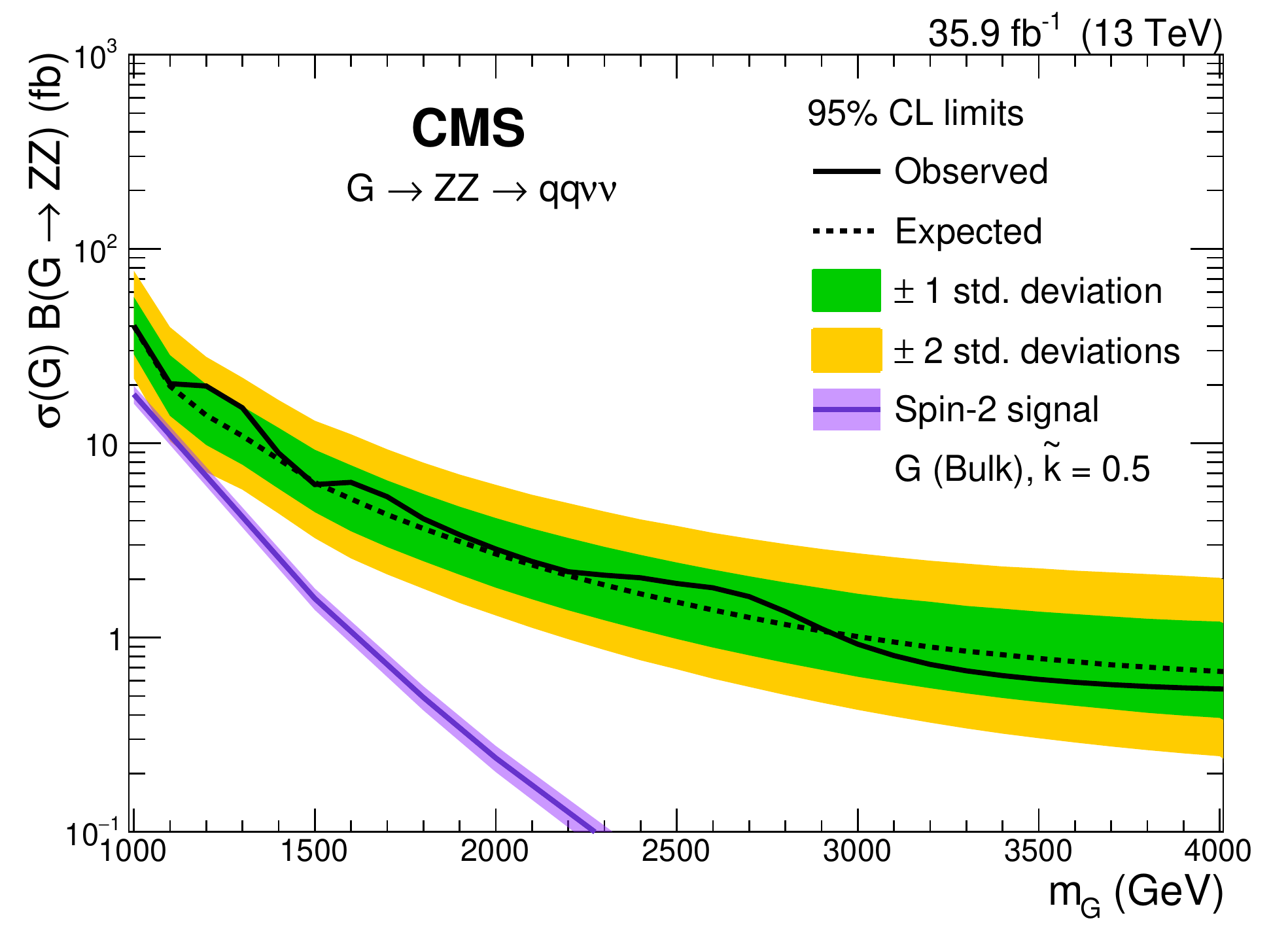}
  \caption{The observed and expected limits on the product of the cross section and branching fraction $\sigma \mathcal{B} (\PWpr \to \PW_{\text{had}} \PZ_{\text{inv}})$ for a spin-1 HVT signal hypothesis (\cmsLeft) and $\sigma \mathcal{B} (\cPG \to \PZ_{\text{had}} \PZ_{\text{inv}})$ for a spin-2 bulk graviton signal hypothesis (\cmsRight), as a function of the \PWpr and {\cPG} mass, respectively. The low- and high-purity categories have been combined. The inner and outer shaded bands indicate the 68\% and 95\% uncertainty intervals associated with the expected limits. Theoretical predictions are shown for: (\cmsLeft) the two HVT models considered, model A (blue dotted-and-dashed line) and model B (red solid line), and (\cmsRight) a graviton model with a curvature parameter of $\ktilde=0.5$ (violet solid line).}
  \label{fig:limitscomb_both}
\end{figure}

The observed and expected limits on the product of the cross section and branching fraction ($\sigma \mathcal{B}(\PWpr \to \PW_{\text{had}} \PZ_{\text{inv}})$) for a spin-1 \PWpr decaying into {\PW} and \PZ bosons that in turn decay in the hadronic and invisible channels, respectively, as a function of the mass of the resonance, are shown in Fig.~\ref{fig:limitscomb_both} (\cmsLeft).
The hypothesis of a heavy spin-1 resonance, predicted by the HVT model A scenario, is rejected at 95\% CL for masses smaller than 3.1\TeV, while the \PWpr described in the HVT model B context is excluded up to 3.4\TeV. At these mass values, the product of cross section and branching fraction are expected to be 1.4\unit{fb} and 1.1\unit{fb}, respectively.

The observed and expected limits on the product of the cross section and branching fraction ($\sigma \mathcal{B}(\cPG \to \PZ_{\text{had}} \PZ_{\text{inv}})$) for a spin-2 bulk graviton decaying into a pair of \PZ bosons, where one \PZ boson decays hadronically and the other invisibly, are shown in Fig.~\ref{fig:limitscomb_both} (\cmsRight), as a function of the mass of the resonance. The theoretical predictions for the curvature parameter hypothesis $\ktilde=0.5$ are shown for comparison.

The results of this search complement those published by the ATLAS Collaboration~\cite{Aaboud:2017itg}, which were obtained from an investigation of the same final state, using different jet substructure and background estimation techniques. The limits obtained here are the best single limits obtained in this final state.

\section{Summary}
\label{Summary}

A search has been made for heavy diboson resonances (\PW\PZ, $\PZ\PZ$) decaying into a pair of vector bosons, one of which is a \PZ boson decaying into $\PGn\PAGn$ and the other is a {\PW} or \PZ boson that decays into $\PQq\PAQq$. The data were collected by the CMS detector from proton-proton collisions produced at the LHC at a center-of-mass energy of 13\TeV. In this analysis, the hadronically decaying {\PW} or \PZ boson is reconstructed as a large-cone jet. The invisible decay of the \PZ boson manifests itself as a large amount of missing transverse momentum recoiling against the jet. The transverse components of the V\PZ system momentum are used to define the transverse mass variable, where a search for a localized excess is performed. The expected background is described with a hybrid data/simulation approach that takes advantage of data sidebands to predict the background normalization and shape in the signal region. To improve the discovery potential, two purity categories are defined, based on a jet substructure observable. An unbinned maximum likelihood fit is performed. No excess is observed in data compared to standard model predictions. Upper limits are established at 95\% confidence level on the product of the production cross section and branching fraction for a spin-1 heavy vector triplet (HVT) \PWpr boson and spin-2 bulk graviton, which are in the range 0.9--63\unit{fb} and 0.5--40\unit{fb}, respectively, depending on the resonance mass. The existence of a \PWpr boson is excluded at 95\% confidence level up to a mass of 3.1\TeV in the HVT model A and up to 3.4\TeV in the HVT model B.

\begin{acknowledgments}
We congratulate our colleagues in the CERN accelerator departments for the excellent performance of the LHC and thank the technical and administrative staffs at CERN and at other CMS institutes for their contributions to the success of the CMS effort. In addition, we gratefully acknowledge the computing centers and personnel of the Worldwide LHC Computing Grid for delivering so effectively the computing infrastructure essential to our analyses. Finally, we acknowledge the enduring support for the construction and operation of the LHC and the CMS detector provided by the following funding agencies: BMWFW and FWF (Austria); FNRS and FWO (Belgium); CNPq, CAPES, FAPERJ, and FAPESP (Brazil); MES (Bulgaria); CERN; CAS, MoST, and NSFC (China); COLCIENCIAS (Colombia); MSES and CSF (Croatia); RPF (Cyprus); SENESCYT (Ecuador); MoER, ERC IUT, and ERDF (Estonia); Academy of Finland, MEC, and HIP (Finland); CEA and CNRS/IN2P3 (France); BMBF, DFG, and HGF (Germany); GSRT (Greece); NKFIA (Hungary); DAE and DST (India); IPM (Iran); SFI (Ireland); INFN (Italy); MSIP and NRF (Republic of Korea); LAS (Lithuania); MOE and UM (Malaysia); BUAP, CINVESTAV, CONACYT, LNS, SEP, and UASLP-FAI (Mexico); MBIE (New Zealand); PAEC (Pakistan); MSHE and NSC (Poland); FCT (Portugal); JINR (Dubna); MON, RosAtom, RAS, RFBR and RAEP (Russia); MESTD (Serbia); SEIDI, CPAN, PCTI and FEDER (Spain); Swiss Funding Agencies (Switzerland); MST (Taipei); ThEPCenter, IPST, STAR, and NSTDA (Thailand); TUBITAK and TAEK (Turkey); NASU and SFFR (Ukraine); STFC (United Kingdom); DOE and NSF (USA).

\hyphenation{Rachada-pisek} Individuals have received support from the Marie-Curie program and the European Research Council and Horizon 2020 Grant, contract No. 675440 (European Union); the Leventis Foundation; the A. P. Sloan Foundation; the Alexander von Humboldt Foundation; the Belgian Federal Science Policy Office; the Fonds pour la Formation \`a la Recherche dans l'Industrie et dans l'Agriculture (FRIA-Belgium); the Agentschap voor Innovatie door Wetenschap en Technologie (IWT-Belgium); the F.R.S.-FNRS and FWO (Belgium) under the ``Excellence of Science - EOS'' - be.h project n. 30820817; the Ministry of Education, Youth and Sports (MEYS) of the Czech Republic; the Lend\"ulet (``Momentum'') Program and the J\'anos Bolyai Research Scholarship of the Hungarian Academy of Sciences, the New National Excellence Program \'UNKP, the NKFIA research grants 123842, 123959, 124845, 124850 and 125105 (Hungary); the Council of Science and Industrial Research, India; the HOMING PLUS program of the Foundation for Polish Science, cofinanced from European Union, Regional Development Fund, the Mobility Plus program of the Ministry of Science and Higher Education, the National Science Center (Poland), contracts Harmonia 2014/14/M/ST2/00428, Opus 2014/13/B/ST2/02543, 2014/15/B/ST2/03998, and 2015/19/B/ST2/02861, Sonata-bis 2012/07/E/ST2/01406; the National Priorities Research Program by Qatar National Research Fund; the Programa Estatal de Fomento de la Investigaci{\'o}n Cient{\'i}fica y T{\'e}cnica de Excelencia Mar\'{\i}a de Maeztu, grant MDM-2015-0509 and the Programa Severo Ochoa del Principado de Asturias; the Thalis and Aristeia programs cofinanced by EU-ESF and the Greek NSRF; the Rachadapisek Sompot Fund for Postdoctoral Fellowship, Chulalongkorn University and the Chulalongkorn Academic into Its 2nd Century Project Advancement Project (Thailand); the Welch Foundation, contract C-1845; and the Weston Havens Foundation (USA).
\end{acknowledgments}

\bibliography{auto_generated}

\cleardoublepage \appendix\section{The CMS Collaboration \label{app:collab}}\begin{sloppypar}\hyphenpenalty=5000\widowpenalty=500\clubpenalty=5000\input{B2G-17-005-authorlist.tex}\end{sloppypar}
\end{document}

%% file: B2G-17-005-authorlist.tex
\vskip\cmsinstskip
\textbf{Yerevan Physics Institute,  Yerevan,  Armenia}\\*[0pt]
A.M.~Sirunyan,  A.~Tumasyan
\vskip\cmsinstskip
\textbf{Institut f\"{u}r Hochenergiephysik,  Wien,  Austria}\\*[0pt]
W.~Adam,  F.~Ambrogi,  E.~Asilar,  T.~Bergauer,  J.~Brandstetter,  E.~Brondolin,  M.~Dragicevic,  J.~Er\"{o},  A.~Escalante Del Valle,  M.~Flechl,  M.~Friedl,  R.~Fr\"{u}hwirth\cmsAuthorMark{1},  V.M.~Ghete,  J.~Grossmann,  J.~Hrubec,  M.~Jeitler\cmsAuthorMark{1},  A.~K\"{o}nig,  N.~Krammer,  I.~Kr\"{a}tschmer,  D.~Liko,  T.~Madlener,  I.~Mikulec,  E.~Pree,  N.~Rad,  H.~Rohringer,  J.~Schieck\cmsAuthorMark{1},  R.~Sch\"{o}fbeck,  M.~Spanring,  D.~Spitzbart,  A.~Taurok,  W.~Waltenberger,  J.~Wittmann,  C.-E.~Wulz\cmsAuthorMark{1},  M.~Zarucki
% ...
\vskip\cmsinstskip
% ...
1:~Also at Vienna University of Technology,  Vienna,  Austria\\
% ...
